\documentclass[a4paper,usenatbib]{mnras} 
\usepackage{lineno,amssymb,graphicx,aas_macros}

\usepackage{graphicx}
\usepackage{graphics}
\usepackage{adjustbox}

\usepackage{mathrsfs}
\usepackage{amsmath}
\usepackage{amssymb}

\usepackage{hyperref}
\usepackage{color}
\usepackage{aas_macros}



\title[The MOND acceleration $a_0$ within SIV]{Elucidating the $z$-dependence of the MOND 
acceleration ($a_0$) within the Scale Invariant Vacuum (SIV) paradigm}

\author[Gueorguiev, V.G.]{
Vesselin G. Gueorguiev $^{1,2}$\thanks{E-mail: Vesselin~at~MailAPS.org; Orcid ID: 0000-0002-2022-6432;}\\
$^{1}${Institute for Advanced Physical Studies, Sofia, Bulgaria}\\
$^{2}${Ronin Institute for Independent Scholarship, Montclair, NJ, USA}
}

\date{Accepted 2024 September 6. Received 2024 July 16; in original form 2024 April 17}

\pubyear{2024}

\begin{document}
\label{firstpage}
\pagerange{\pageref{firstpage}--\pageref{lastpage}}
\maketitle

\begin{abstract}
In a recent paper: ‘‘On the time dependency of $a_0$''  
the authors claim that they have tested 
``one of the predictions of the Scale Invariant Vacuum (SIV) theory on MOND''
by studying the dependence of the Modified Newtonian Dynamics (MOND) 
acceleration at two data sets, low-$z$ ($3.2\times10^{-4}\le z\le 3.2\times10^{-2}$) and high-$z$  ($0.5\le  z\le  2.5$).
They claim ``both samples show a dependency of $a_0$ from $z$''. 
Here, the work mentioned above is revisited.
The explicit analytic expression for the $z$-dependence of the 
$a_0$  within the SIV theory is given.
Furthermore, the first estimates of the $\Omega_m$ within SIV theory give $\Omega_{m}=0.28\pm0.04$ 
using the low-z data only, while a value of  $\Omega_{m}=0.055$ is obtained using both data sets. 
This much lower $\Omega_m$ leaves no room for non-baryonic matter! Unlike in 
the mentioned paper above, 
the slope in the $z$-dependence of $A_0=\log_{10}(a_0)$ 
is estimated to be consistent with zero Z-slope for the two data sets. Finally, 
the statistics of the data are consistent with the SIV predictions; 
in particular, the possibility of change in the sign of the slopes for the two data sets
is explainable within the SIV paradigm; 
however, the uncertainty in the data is too big for the clear 
demonstration of a $z$-dependence yet.
\end{abstract}
\begin{keywords}
cosmology -- dark matter -- cosmological parameters -- cosmology: theory
\end{keywords}


\section{Introduction} \label{sec:intro}

Modern physics is well understood based on two main contemporary pillars:
Einstein's General Relativity (EGR) and the Relativistic Quantum Field Theory. 
However, there are some perplexing observations about the 
motion of stars  within galaxies and clusters. 
Within the popular current model of Cosmology and Astrophysics, 
the resolution of these perplexing phenomena is often associated with 
concepts such as Dark Matter and Dark Energy  \cite{DM-History-Bertone'18}.
However, for over 30 years, there has not been a definite detection of any new particles or fields.
An alternative to the Dark Matter approach to resolving the observational discrepancies
in galaxies and clusters of galaxies is the idea of the Modified Newtonian Dynamics (MOND, \cite{MOND}) 
that has steadily gained support in the Astrophysics communities. 
While the concept of  Dark Matter is a natural continuation of the matter paradigm into 
a non-luminous matter to explain the observational fact of the flat rotational curves, 
the MOND idea does not need extra matter\footnote{Some 
fashionable models of gravity are trying to explain MOND and its
fundamental acceleration $a_0$ \citep{entropic_gravity_and_MOND,Emergent_Gravity},
but these models have failed the observational tests by \cite{Testing_Verlinde's_EG},
while the relativistic implementation of MOND 
suggests imperceptible variation of $a_0$ ``to redshift unity or even beyond it''
\citep{varing_G_and_a0}.}; instead, it modifies the dynamics 
once the observed acceleration $g=v^2/r$ falls below the certain cut-off value $a_0\gg\,g$.
In this deep  MOND regime, one expects scale invariance 
to be present in the system under study \citep{MOND_and_ScaleInv'09}. 

Scale invariance is an old idea introduced by Weyl as early as 1918 
\cite{Weyl-Geometry-Weyl'22-English, Weyl-Geometry-Weyl'18-ed6}
as a gauge invariant gravity, where along with the metric tensor $g_{\mu\nu}$
there is  a connexion vector $\kappa_\mu$ controlling the length change $dl=l\kappa_\mu\,dx^\mu$,
and a scalar field $\lambda$  that describes the gauge freedom  $g_{\mu\nu}\rightarrow\lambda^2g_{\mu\nu}$.
The shortcomings of the original Weyl geometry pointed out by \cite{Weyl-Geometry-Objection-Einstein'18} 
were addressed by the introduction of the 
Weyl Integrable Geometry (WIG) \citep{Eddington'23},
where the connexion vector satisfies $\kappa_\mu=-\partial_\mu\ln{\lambda}$.
Consequently, \cite{Weyl-Geometry-Dirac'73} and \cite{Scale-covariance-Canuto'77} 
have applied the idea to formulate scale invariant cosmology
and tried to fix $\lambda$ based on Dirac's Large Numbers Hypothesis \citep{Dirac-LargeNumbers'74}.
The recent reincarnation of the notion of scale invariance was introduced by \citet{Maeder'17a,Maeder'17c},
where the scalar field $\lambda$ was fixed to be only time-dependent 
by the requirement of homogeneity and isotropy of space. 
In doing so, the specific functional form of  $\lambda(t)$ is
determined by the requirement that the macroscopic vacuum must be scale invariant 
and thus introducing the Scale Invariant Vacuum (SIV) paradigm \citep{SIV-PartI'23}.
This new approach has been explored only in the past few years by \cite{SIV_Results'24}
as a potential alternative to the standard cosmological 
model of dark energy plus cold dark matter 
paradigm ($\Lambda$CDM) \cite{CosmolParameters-Planck'15results}.
This new alternative approach suggests a possible connection to dark matter and dark energy \citep{SIVandDM}.
It has  been shown recently by \citet{SIVandMOND} that the MOND fundamental acceleration $a_0$ 
could be derived within the SIV-paradigm, and the result depends on the cosmological parameters
such as Hubble constant $H_0$ and the total current mass fraction $\Omega_m=\rho_m/\rho_c$,
where $\rho_c=3H_0^2/(8\pi\,G)$ is the critical density.

By taking this result at face value along with the epoch-dependent scale factor $\lambda(t)$, 
it is natural to expect that the SIV-derived MOND acceleration $a_0$ 
may have an epoch-dependent value, 
just as it is the case for the mass content of the Universe $\rho_m$, 
and the Hubble parameter  as well ($H=\dot{a}/a$, 
where $a(t)$ is the usual FLRW expansion factor).

In this respect, the recent papers by \citet{Popolo'24a,Popolo'24c} have initiated interesting research 
about testing the connection between MOND by \cite{MOND} and 
its possible justification within the SIV paradigm by \cite{SIVandMOND}.
In doing so, they studied the $z$-dependence of $a_0$ using observational data
but didn't derive the explicit $z$-dependence, 
nor did they discuss the relevant SIV model parameters for $\Omega_m$.
As a new model different from $\Lambda$CDM, 
one should expect that some of the standard cosmological parameters 
may have different values within the SIV model. In this case, 
$\Omega_m$ is a model parameter to be determined, while the Hubble constant $H_0$ 
is  a model-constraining observational parameter. 

In what follows, I will present my analyses of the
$z$-dependency of the MOND acceleration $a_0$
along with the specific $z$-dependent expression of $a_0$ within SIV.
Furthermore, the results of the statistical analyses will be utilized to 
perform one of the first determinations of the SIV parameter $\Omega_m$
representing the fraction of the total matter-energy content of the Universe.
The results will illustrate a puzzling situation that needs a better understanding
of the data or the model utilized.

\section{Results from Statistics} \label{sec:results}

\begin{table}
\begin{center}
\caption{\label{Table1} Results from simple statistical analysis for the two sets of redshifts; 
the first set of low-$z$ data with $0.00032 \le z \le 0.032$ \citep{Marra'20}
and the second high-$z$ data with $0.5 \le  z \le  2.5$ \citep{Popolo'24c}.}
\begin{tabular}{c|cc}
data &  $\bar{Z}$ &  $\bar{A_0}$ \\
\hline
low-$z$ & $-2.49\pm0.04$ & $-13.07\pm0.06$ \\
high-$z$ & $+0.17\pm0.02$ & $-12.60\pm0.03$ \\
\end{tabular}
\end{center}
\end{table}%

Before going into more detail about the SIV theory, it is important to note 
that a simple statistical analysis of the two main variables $A_0=\log_{10}(a_0)$
with $a_0$ in km/s$^2$ and $Z=\log_{10}(z)$ based on data reported by \citet{Popolo'24c} 
gives averaged values  $\bar{A_0}=-13.07\pm0.06$ and $\bar{Z}=-2.49\pm0.04$ (see Table \ref{Table1})
for the low-$z$ data set from 
\cite{Marra'20}.

For notational and pragmatic reasons, I maintain the choice of the main variables 
to be the dimensionless $A_0$ and $Z$. 
Using, the $\log_{10}$ on $a_0$ in km/s$^2$ keeps the corresponding range of $A_0$
between $-15$ and $-10$, while the  $\log_{10}$ on $z$ is in the range $-4$ to $1$.
Since $a_0$ should be in km/s$^2$ for evaluating $\log_{10}(a_0)$ to compute $A_0$,
that is, $A_0=\log_{10}(a_0/(\textrm{km/s}^2))$ for $a_0$ in arbitrary units, then
the units of $A_0$ are dimensionless, and so are the corresponding average values $\bar{A_0}$.

\begin{figure*}
\centering
\begin{adjustbox}{center}
\includegraphics{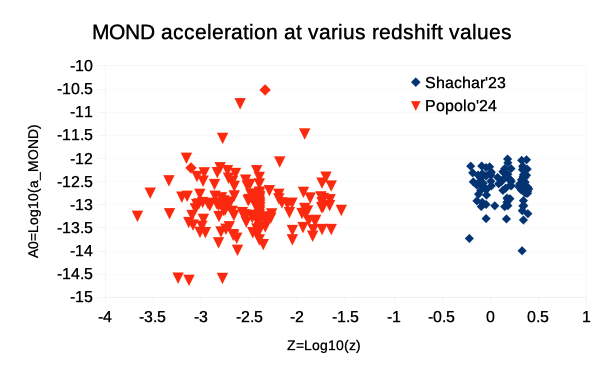}
\end{adjustbox}
\caption{Low-$z$ and high-$z$ data sets are discussed in the text.
For this figure only, I have dropped the outlier related to the NGC2976 Galaxy data in the low-$z$ data set 
since its value $A_0=-18.32$ is too low compared to the other values shown; 
however, the data point has been used in the various data analyses presented.
Unlike in \citet{Popolo'24c}, where only 17 high-$z$ data points have been used, 
here I have shown all 100 data points related to the data in \citet{Shachar'23}.
For the relevant error bars, the reader is referred to the paper by \citet{Popolo'24c}.}
\label{A0(Z)}
\end{figure*}

The high-$z$ data shown in \citet{Popolo'24c} is based on the work done by \citet{Shachar'23}.
It contains only 17 high-$z$ Galaxies for which the inferred $a_0$ is less than 
$1.2\times 10^{-13}$~km/s$^2$; that is, $A_0<-12.92$. 
Such selection criteria cut a large data segment  from the high-$z$ data, 
while it has not been applied to the low-$z$ data, thus introducing a bias.  
In my opinion, a fair,  unbiased, and appropriate data selection procedure, 
if any, should be applied to both sets. Here, I disagree
with such a selection criteria applied to high-$z$ data and not to the  low-$z$ data,
so I use all 100 data points in \citet{Shachar'23} as seen in Fig.  \ref{A0(Z)}, 
both data sets have a compatible spread of $A_0$-values.
The relevant values follow the calculation of $a_0$ using Eq. (5) in \citet{Popolo'24c}
based on the data from \citet{Shachar'23}. 
In doing the calculations, an error was noticed in the initial evaluations by \citet{Popolo'24a} 
due to units conversion; upon communicating with the authors, this was later recognized by \citet{Popolo'24b} 
and corrected in the subsequent version of their paper  \citet{Popolo'24c}; 
the proper conversion is necessary to obtain  $a_0$ in km/s$^2$ for the correct evaluation of the corresponding $A_0$.
It is worth noticing that  the corresponding simple statistical analysis gives
$\bar{A_0}=-12.60\pm0.03$ and $\bar{Z}=0.17\pm0.02$ 
for all of the 100 high-$z$ data points (see Table \ref{Table1}). 
Thus, the $Z$-slope based on these two aggregated data sets is 
$\Delta{\bar{A_0}}/\Delta{\bar{Z}}=(-12.6+13.07)/(0.17+2.49)\approx 0.18\pm0.03$
indicating a change in the MOND acceleration as seen from  Table \ref{Table1}.
This is an overly simplified estimate of the $Z$-slop based on both sets, 
marked in Table \ref{Table2} with an asterisk.


\begin{table}
\small
\caption{\label{Table2}
Values of the Z-slopes and intercepts for the current work and the values deduced in ref. \citep{Popolo'24c}.}
\begin{center}
\begin{tabular}{|c|cc|cc|}
\hline
  &  Current  & work & Del Popolo  & \& Chan  \\
\hline
data & intercept & Z-slope & intercept & Z-slope \\
\hline 
low-$z$  & $-12.8\pm0.3$  &   $0.12\pm0.13$  &$-11.6\pm0.5$ & $+0.6\pm0.2$\\
high-$z$ & $-12.60\pm0.05$ &  $0.01\pm0.2$ & $-9.6\pm0.1$ & $-0.2\pm0.4$\\
both &  $-12.6\pm0.1$ & $0.18\pm0.03^*$  &$-8.6\pm0.5$ & $+0.6\pm0.2$\\
\hline 
\end{tabular}
\end{center}
\end{table}%

The statistical analyses can be taken further, 
as done in the paper by \citet{Popolo'24c}, where they do a linear fit to the two sets and derive
a $Z$-slope of $0.6\pm0.2$ with an intercept of $A_0=-11.6\pm0.5$ for the low-$z$ data. 
Simple linear regression on the same data gives agreement with the zero slope since   
the result is a $Z$-slope of $0.12\pm0.13$ with a different intercept of $A_0=-12.8\pm0.3$.
Such intercept should be regarded as related to $Z=0$; 
thus, to the value of $a_0$ at $z$=1 and not at $z=0$. 
Therefore, this corresponds to about $11\%$ or only $2\%$
change of $A_0$ from $\bar{A_0}=-13.07$ near $z\approx0$ 
to  $A_0=-11.6$ or $A_0=-12.8$ near $z\approx1$.

Regarding the high-$z$ data, the $Z$-slope of $-0.2\pm0.4$ is consistent with zero 
while the intercept of $A_0=-9.61\pm0.08$ for the high-$z$ data is questionable
due to the applied data selection criteria and the possible error of their analyses mentioned earlier. 
The simple linear regression on the full data set derived from \citet{Shachar'23}
gives again agreement with the zero $Z$-slope since   
the result is $Z$-slope is $0.01\pm0.2$ with an intercept of $A_0=-12.603\pm0.05$. 
Notice that for the current analysis, the low-$z$ and high-$z$ 
data sets have intercepts that agree with each other unlike those in \citet{Popolo'24c}. 
{\it There must be an agreement between these two
intercepts since they should reflect the value of the MOND acceleration at $z=1$.} 

\section{SIV framework} \label{sec: framework}

Within SIV the fundamental MOND acceleration $a_0$ can be related to the Hubble constant $H_0$ 
and the current matter content of the Universe $\Omega_m$ \citet{SIVandMOND, SIVandDS}. 
For the derivations and formulas to be used in this section, 
I will denote the MOND fundamental acceleration $a_0$ by $a_{M}$ whenever appropriate
to avoid confusion with the expansion scale factor $a$ but will use $a_0$ in the absence of such a problem.
This is done to avoid confusion with the FLRW expansion scale factor $a$, 
which by convention should be denoted by $a_0$ at the current epoch; 
and to also avoid awkward notation $a_{00}$ for the current value of the MOND acceleration.

Within SIV, there is an extra velocity-dependent term,
denoted as dynamical acceleration \cite{SIVandDM}: 
\begin{equation}
\frac{d^{2}\overrightarrow{r}}{dt^{2}}
=-\frac{G_{t}M(t)}{r^{2}}\frac{\overrightarrow{r}}{r}+\kappa(t)\frac{d\overrightarrow{r}}{dt},
\label{eq: SIV dynamics}
\end{equation}
where $\kappa=-\dot{\lambda}/\lambda$ is the time component of the SIV connexion vector, 
with a simple functional form $\kappa=1/t$ within the SIV gauge,
where the SIV cosmic time $t$ is a dimensionless parameter
such that $t\in[t_{in},t_0=1]$. Here, $t_{in}$ is the moment of the Bing Bang when 
the FLRW scale factor satisfies $a(t_{in})=0$, happening near $t_{in}=\Omega^{1/3}$,
while at the current epoch $a(t_{0})=1$ and time is set so that $t_{0}=1$. 
Within SIV the conformal-scale factor is $\lambda=1/t$ and is
used to perform Weyl transformation  $g'_{\mu\nu}=\lambda^2 g_{\mu\nu}$ 
that relates EGR metric $g'_{\mu\nu}$ to the 
metric $g_{\mu\nu}$ within the WIG framework \citep{SIVandDM}.    

To arrive at an expression for the MOND acceleration $a_{0}$, 
one considers the ratio of  the magnitudes of 
the Newtonian acceleration $g_N = GM/r^2$ 
to the additional acceleration $\kappa(t)v$ in \eqref{eq: SIV dynamics}, 
where $v$ denotes the magnitude of the 
velocity $\overrightarrow{v}=d\overrightarrow{r}/dt$:
\begin{equation}
x=\frac{\kappa vr^{2}}{GM}.\label{eq: x}
\end{equation}
Now, one can use the relation given by the instantaneous radial acceleration
$v^{2}/r=GM/r^{2}$ to eliminate the speed $v$ from the expression of $x$ given by \eqref{eq: x}; 
then using $g_{N}=GM/r^{2}$ to remove $GM$, one arrives at:
\begin{equation}
x=\frac{\kappa vr^{2}}{GM}=\kappa\sqrt{\frac{r^{3}}{GM}}=\kappa\sqrt{\frac{r}{g_{N}}}.
\end{equation}

When the dynamic acceleration dominates over the Newtonian acceleration ($x \gg 1$), one has:
\begin{equation}
g=g_{N}+xg_{N}\approx xg_{N}=\kappa\sqrt{rg_{N}}.
\end{equation}
Therefore, one arrives at the MOND type relation $g\sim\sqrt{a_0g_{N}}$
from which one can deduce an expression for $a_0$:
\begin{equation}
a_{0}\approx\kappa^{2}r.\label{eq:a_0(r)}
\end{equation}
The above expression indicates a possible $r$ dependence of $a_0$,
thus demoting $a_0$ of its fundamental parameter status within MOND,
which may be testable with future high-precision data. 
Such dependence may explain the variance of $a_0$.
To restore the fundamental character of 
the above expression \eqref{eq:a_0(r)} as in MOND,
one could consider the limit $r\rightarrow\,r_H$,
where the Hubble radius reflects the influence of the Universe
causally connected to the object studied.
Thus, the time-dependent MOND acceleration within SIV
is the upper bound of \eqref{eq:a_0(r)} given by
the expression:
\begin{equation}
\kappa^{2}r\rightarrow \kappa^{2}r_H=\kappa^{2}c/H=a_{M}(t),\label{eq:a_0 within SIV}
\end{equation}

During the matter-dominated epoch, SIV has an analytic form for expansion scale-factor $a(t)$ 
\citep{Jesus'17,SIVandDM}:
\begin{equation}
a(t)=\left(\frac{t^{3}-\Omega_{m}}{1-\Omega_{m}}\right)^{2/3}
\Rightarrow H= \frac{2t^{2}}{t^{3}-\Omega_{m}}.\label{eq:a(t) and H(t)}
\end{equation}

\noindent
Thus, from (\ref{eq:a_0 within SIV}) with SIV time in units $t\in[t_{in},\:t_{0}=1]$, 
one obtains for the MOND fundamental acceleration:
\begin{equation}
a_{M}(t)=c\frac{(t^{3}-\Omega_{m})}{2t^{4}}.\label{eq:a_0(t)}
\end{equation}
To express $a_{M}$ in the usual time units, where $\tau_{in}=0$
at the Big Bang when the scale factor $a=0$ while the age of the
Universe now is $\tau_{0}=13.8$ billion years, one has to use the
chain rule for differentiation, that is:

\begin{equation}
a_{M}(\tau)=c\left(\frac{\dot{\lambda}}{\lambda}\right)^{2}\frac{a}{\dot{a}}=
c\kappa(\tau)\frac{\kappa(t)}{H(t)}=\left(\frac{dt}{d\tau}\right)a_{M}(t).
\label{eq:a_0(tau)}
\end{equation}

The value of $dt/d\tau$ is assessed based on the assumption that
the following relation provides a connection between the two time-scales 
$(t-t_{in})/(t_{0}-t_{in})=(\tau-\tau_{in})/(\tau_{0}-\tau_{in})$
where $t_{in}=\Omega_{m}^{1/3},\:t_{0}=1,\:\tau_{in}=0,$ and $\tau_{0}=13.8$
billion years \citep{SIVandDM}. Thus, one has: 
\begin{equation}
dt/d\tau=(1-t_{in})/\tau_{0},\label{eq:dt/dtau}
\end{equation}
and therefore: 
\begin{eqnarray}
a_{M}(\tau)&=&c\left(\frac{1-\Omega_{m}^{1/3}}{\tau_{0}}\right)
\left(\frac{t^{3}-\Omega_{m}}{2t^{4}}\right)=\\
&=&
\frac{c}{2b}\left(\frac{1-b}{\tau_{0}}\right)\left(\frac{x^{3}-1}{x^{4}}\right),
\label{eq:a_0(x)}
\end{eqnarray}
where $x=t/b$, $\Omega_{m}=t_{in}^{3}=b^{3}$. 
Set $\widetilde{a}=a\left(\frac{1-b^{3}}{b^{3}}\right)^{2/3}$ 
and revisit (\ref{eq:a(t) and H(t)}) to get:
\begin{equation}
a=\left(\frac{b^{3}}{1-b^{3}}\right)^{2/3}\left(x^{3}-1\right)^{2/3}
\Rightarrow\widetilde{a}=\left(x^{3}-1\right)^{2/3}.
\end{equation}
Thus, upon utilization of \eqref{eq:a_0(x)} and the above substitution,  
the MOND acceleration (\ref{eq:a_0(tau)}), as a function of the scale factor, 
becomes:
\begin{equation}
a_{M}(\tau)=\frac{c}{2b}\left(\frac{1-b}{\tau_{0}}\right)
\frac{\widetilde{a}^{3/2}}{\left(\tilde{a}^{3/2}+1\right)^{4/3}}.\label{eq:a_0(a-tilda)}
\end{equation}
Next, look at the $z$-dependence, and use $a=1/(z+1)$; 
thus, when $z$ is $0$, or $1$, and even $2$, 
then $a$ is $1$, or $1/2$, and correspondingly $1/3$. 
Therefore, one has: 
\begin{eqnarray}
a_{M}(z)&=&a_{M}(z=0)\times
\frac{\left(\frac{1}{z+1}\right)^{3/2}\left(\tilde{a}_{0}^{3/2}+1\right)^{4/3}}
{\left(\left(\frac{1}{z+1}\right)^{3/2}\tilde{a}_{0}^{3/2}+1\right)^{4/3}}\\
&=&a_{M0}\times\left({z+1}\right)^{-3/2}
\left(\frac{\left(1-\Omega_{m}\right)}{\left({z+1}\right)^{3/2}}+\Omega_{m}\right)^{-4/3}
\label{eq:a_0(z)}
\end{eqnarray}
where $\tilde{a}_{0}=\left(\frac{1-\Omega_{m}}{\Omega_{m}}\right)^{2/3}$ was utilized.
The above expression \eqref{eq:a_0(z)} provides the formula for the explicit 
$z$-dependence of the MOND fundamental acceleration 
within the SIV framework. It can be used to
test this SIV prediction against the observational data. 
When evaluated at $z=1$ and $2$, one finds that $a_{M}(z)$
is about $79\%$ and $58\%$ of the current value $a_{M0}=a_{M}(z=0)$
for $\Omega_{m}=0.3$. Thus, the $z$-dependence of the MOND acceleration
is weak, and it is likely buried within the scatter of the current
observational data and its uncertainty (see \citet{Popolo'24c}); 
as one can see later, the value of $a_{M0}$, at best, is accurate within 13\%
while the $1\sigma$ error of the data on $A_0$ reported by \citet{Popolo'24c}
usually translates in more than $29\%$ uncertainty for observationally deduced 
MOND acceleration data points.

The value of $a_{M0}$ could be used
to assess the parameter $\Omega_{m}$ within the SIV theory. 
That is, by using (\ref{eq:a_0(x)}) one has:
\begin{equation}
a_{M0}=\frac{c\,(1-b)(1-b^{3})}{2\tau_{0}}=
\frac{c(1-\Omega_{m}^{1/3})(1-\Omega_{m})}{2\tau_{0}}.\label{eq:a_M0 and Omega_m}
\end{equation}
One can solve for $\Omega_{m}$ by taking $\log_{10}$ of (\ref{eq:a_M0 and Omega_m}):
\begin{equation}
A_{0}-\log_{10}(\frac{c}{2\tau_{0}})=
\log_{10}\left[\left(1-\Omega_{m}^{1/3}\right)\left(1-\Omega_{m}\right)\right],
\label{eq:A0}
\end{equation}
for low-$z$ data with $\bar{A_{0}}=-13.07$ this gives $\Omega_{m}=0.28$.
Such a value of $\Omega_{m}$ aligns with the comparative study of the SIV and 
$\Lambda$CDM cosmologies that demonstrated only light adjustments in 
$\Omega_{m}$ if the expansion factors $a(t)$ 
of these models are to be very similar \cite{Maeder'17a}. 
However, it contrasts with the MOND idea of dark matter redundancy.
Notably, the range of values of $A_{0}$ based on \eqref{eq:A0} for 
$\Omega_{m}$ given by $0.01, 0.1, 0.3, 0.6, 0.8$ are
$-12.57, -12.78, -13.098, -13.67, -14.31$.

Based on (\ref{eq:a_M0 and Omega_m}), 
the corresponding fractional uncertainties are related in the following way:
\begin{equation}
\frac{\Delta a_{M0}}{a_{M0}}=\frac{\Delta\Omega_{m}}{\Omega_{m}}
\left(\frac{\Omega_{m}}{\left(1-\Omega_{m}\right)}+\frac{1}{3}
\frac{\Omega_{m}^{1/3}}{\left(1-\Omega_{m}^{1/3}\right)}\right).
\label{eq:fractinal error}
\end{equation}
The slope of the $z-$dependence, $m=\frac{d\log_{10}a_{M}(z)}{dz}$
when taking $\log_{10}$ of (\ref{eq:a_0(z)}), is then:
\begin{equation}
m=\frac{1-(1+3(1+z)^{3/2})\Omega_{m}}{(1+z)(1+((1+z)^{3/2}-1)\Omega_{m})\ln(100)}.
\label{eq:z-slope}
\end{equation}
giving $m_0=(1-4\Omega_m)/\ln(100)$ at $z=0$, 
which is positive only for $\Omega_m<0.25$.
While for the $Z$-slope one has:
\begin{equation}
\frac{dA_{0}}{dZ}=\left(\frac{dA_{0}}{dz}\right)\left(\frac{dz}{dZ}\right)=
m\ln(10)z=m\ln(10)10^{Z},
\end{equation}
where $dz/dZ=(\exp(Z\ln(10)))'$ is utilized.
Based on the data provided: $A_{0}=\log_{10}a_{M0}=-13.067\pm0.056$
for $a_{M0}$ in km/s$^{2}$. The fractional uncertainty is
$\Delta A_{0}/|A_{0}|=0.0043=\Delta(a_{M0})/(|A_{0}|a_{M0}\ln(10))$.
That is, $\Delta(a_{M0})/a_{M0}=0.13$; therefore,
using (\ref{eq:fractinal error}) one has $\Omega_{m}=0.28\pm0.04$.
The $z$-slope is then $m_0=-0.01\pm0.3$, 
while the $Z$-slope is $-0.0001$ using $\overline{Z}=-2.49$. 
This results practically in a horizontal line that changes very little 
from being $-13.0667$ at $Z=-4$ to $-13.0671$ at $Z=2$. 
This change is well within the current error ($\pm0.06$) for the low-$z$.
For the high-$z$ data, one can now evaluate  $m_1$ to be $-0.11$;
therefore, the $Z$-slope is $-0.25$ in agreement with 
the $Z$-slope of $-0.2\pm0.4$ reported by \cite{Popolo'24c}.

\section{Discussion and Conclusion} \label{sec: conclusion}

It is still inconclusive about the $z$-dependence
of the MOND fundamental acceleration, 
but such dependence is present within the SIV theory \eqref{eq:a_0(z)};
furthermore, the SIV expression (\ref{eq:z-slope}) does suggest 
that there is a change in the sign of the slope when going through 
$z=\left(1/3\times(1-\Omega_m)/\Omega_m\right)^{2/3}-1$
with $m_0$ positive for $\Omega_m<0.25$ and negative otherwise;
this could explain the change in the sign of the slope of the 
two data sets as noticed by \citet{Popolo'24c}.
However, the corresponding value of $A_0$ at $z=1$,
which is the intercept, is about the same as $\bar{A_0}$
based on low-z with $\bar{A_0}=-13.07$, 
but Del Popolo \& Chan's high-$z$ intercept differs significantly from the 
corresponding $\bar{A_0}$ values around  $A_0=-12.6$.
However, if one is to embrace the matching values of $A_0$ at $z=1$, that is, 
to use  $A_0=-12.6\pm0.05$ along with (\ref{eq:a_M0 and Omega_m}) in (\ref{eq:a_0(z)}), 
Then, the value of $\Omega_m$ is significantly lower; 
that is, one gets $\Omega_m=0.055$.
In this case, the sign change of the slopes will be happening around $z=2.7$.
For $\Omega_m=0.055$, equation \eqref{eq:A0} gives $A_0=-12.69$, 
which is lower than $A_0=-12.6$ but is bigger than $A_0=-13.07$.
Note that such a low value for $\Omega_m$ does not leave much room for any dark matter.
Such a result is better aligned with the MOND view about dark matter but seems to be 
a drastic departure from the need for dark matter and dark energy as required by $\Lambda$CDM.
This approach may be favorable as a method of determining $\Omega_m$ since it relies on 
the two data sets and their consistent $Z$-intercepts, 
which is in contrast to the first method presented 
that utilized only the low-$z$ data
and assumed that $\bar{A_0}=-13.07$ at $\bar{Z}=-2.49$ is sufficiently close to $z=0$
even though $z\rightarrow0$ would imply $Z\rightarrow -\infty$.
Therefore, further studies are needed to determine the correct $\Omega_m$ values within the SIV framework. 
Thus, more precise data analyses are needed along with improved uncertainties of the 
observational data points (for the currently used data see Fig. \ref{A0(Z)}) 
to confirm the $z$-dependence of the 
MOND fundamental acceleration and to potentially test the  SIV theory via 
its model prediction for $a_0(z)$ as well as to deduce the relevant SIV model parameters.

In conclusion, the long-standing mystery of galactic rotation curves has 
fueled the development of Modified Newtonian Dynamics (MOND). 
This work presents a significant contribution by providing the first explicit analytic expressions 
for the z-dependence of the fundamental MOND acceleration ($a_0$) 
within the framework of the Scale Invariant Vacuum (SIV) theory  \eqref{eq:a_0(z)}. 
This novel approach goes beyond previous studies \citet{Popolo'24a,Popolo'24c}.
Furthermore, we leverage existing observational data to perform the first-ever 
estimation of the cosmological matter density parameter ($\Omega_m$) within the SIV framework. 
The current analysis yields a value of $\Omega_{m}=0.28\pm0.04$ based on low-z data and
 $\Omega_{m}=0.055$ based on the consistency of both data sets at $z=1$,
potentially removing the need for dark matter entirely. 
The above is a puzzling result as to why the two methods presented to determine the value of $\Omega_{m}$
within SIV result in relevant values within the $\Lambda$CDM model.

On the one hand, the SIV value $28\%$ for $\Omega_{m}$ deduced by using only the dataset with $z\approx0$ 
is close to the $\Lambda$CDM model of about $30\%$ \citep{CosmolParameters-Planck'15results}, 
while on the other hand, the value $5.5\%$ deduced by using both datasets (via the $A_0$ intercept at $z=1$) 
is close to the baryon matter value within the $\Lambda$CDM model of about $5\%$. 
This could be just a numerical coincidence,
or there may be some deeper reason for why the values are like that. 
For example, it may be related to the transition from a matter-dominated epoch 
to a cosmological constant (dark-energy) dominated epoch within the $\Lambda$CDM model. 
However, within the SIV paradigm, one does not expect dark-matter and dark-energy components.
For example, the energy density $\Omega_{\Lambda_E}$ 
due to the Einstein Cosmological Constant $\Lambda_E$,
which within the $\Lambda$CDM model is estimated to be $70\%$,
does not exist within SIV but is replaced by
$\Omega_\lambda=-\frac{2}{H}\frac{\dot{\lambda}}{\lambda}$
that also compliments $\Omega_m$ to 1 (assuming flat Universe $\Omega_k=0$)  
within the SIV paradigm \citep{SIVandDM}.

Interestingly, the data suggests an almost flat $z$-dependence of $A_0=\log_{10}(a_0)$, 
contrasting with previous claims by \citet{Popolo'24a,Popolo'24c}.
While the current data limitations prevent the definitive confirmation of the $z$-dependence \eqref{eq:a_0(z)}, 
the observed trends are consistent with SIV predictions.  
SIV offers a unique explanation for the potential sign change in the 
slopes previously indicated across different redshift ranges.
Future higher precision data will be crucial for definitively 
resolving the presence or absence of z-dependence in $a_0$.

\section*{Acknowledgments}
The author is grateful 
for the moral and financial support by particularly close private parties
during the various stages of the research presented 
and to Prof. A. Maeder for the long and fruitful scientific collaboration over the years. 
This research does not receive any specific grant from funding agencies in the public, 
commercial, or not-for-profit sectors. 
However, the publishing charges were waived graciously   
by the Author Support, Oxford Journals, at Oxford University Press -
for which the author is very grateful!

\section*{Data availability}
No new data was generated or analyzed in support of this research.



\bibliographystyle{mnras}
\bibliography{DARK43_2024_101393}



\bsp	
\label{lastpage}
\end{document}